# Title: Polychromatic soliton molecules

**Authors:** Joshua P. Lourdesamy[1], Antoine F. J. Runge[1]*, Tristram J. Alexander[1], Darren D. Hudson[2], Andrea Blanco-Redondo[3], C. Martijn de Sterke[1,4]

**Affiliations:**
[1]Institute of Photonics and Optical Sciences (IPOS), The University of Sydney, NSW 2006, Australia.
[2]CACI Photonics Solutions, 15 Vreeland Road, Florham Park, NJ 07932, USA.
[3]Nokia Bell Labs, 791 Holmdel Road, Holmdel, NJ 07733, USA.
[4]The University of Sydney Nano Institute (Sydney Nano), The University of Sydney, NSW 2006, Australia.
*Corresponding author. Email: antoine.runge@sydney.edu.au

**Abstract:** Soliton molecules, bound states composed of interacting fundamental solitons, exhibit remarkable resemblance with chemical compounds and phenomena in quantum mechanics. Whereas optical molecules composed of two or more temporally locked solitons have been observed in a variety of platforms, soliton molecules formed by bound solitons at different frequencies have only recently been theoretically proposed. Here, we report the observation of polychromatic soliton molecules within a mode-locked laser cavity, achieving the desired dispersion by implementing a spectral pulse-shaper. This system supports two or more coincident solitons with different frequencies, but common group velocities. Our results open new directions of exploration in nonlinear dynamics within systems with complex dispersion and offer a convenient platform for optical analogies to mutual trapping and spectral tunneling in quantum mechanics.

Solitons, solitary waves exhibiting particle-like behavior, are ubiquitous in numerous nonlinear physical systems, including waves on a water surface, plasmas, sound waves in superfluids, Bose-Einstein condensates and ultrafast lasers (*1–4*). In optics, soliton effects have been used to generate ultrashort pulses from solid-state and fiber laser cavities (*5*, *6*) and are fundamental building blocks of complex and very significant phenomena such as supercontinuum generation (*7*) and frequency combs (*8*, *9*). Conventional optical solitons arise from the balance between self-phase modulation (SPM) and negative quadratic dispersion. Higher order dispersive effects have historically only been studied as perturbations of conventional temporal solitons, limiting the pulse duration, energy achievable or even affecting the stability of mode-locked lasers (*10–14*).

This paradigm changed with the discovery of pure-quartic solitons (PQSs), a novel class of shape-maintaining pulses (*15*). In the formation of PQSs, the SPM is counterbalanced by negative quartic ($\beta_4 < 0$) dispersion. PQSs display distinctly different characteristics from conventional solitons (*16*), some of which are of interest for application in microresonator-based frequency combs and ultrafast lasers (*17–19*). Despite these differences, numerical studies have recently shown that conventional solitons and PQSs are special cases of a broader family of Generalized Dispersion Kerr Solitons (GDKSs), arising from the interplay between Kerr nonlinearity and a combination of quadratic and quartic dispersion (*20*, *21*).

Amongst this vast family, Tam *et al.* numerically discovered a set of intriguing members which form in the presence of positive quadratic and negative quartic dispersion (*20*). These pulses consist of two solitons centered at two different frequencies, with identical group-velocities, which form a bound state. The associated temporal shape consists of a fast carrier modulated by a meta-envelope, so called because it modulates a function which itself is an envelope (*20*). Other numerical studies reported similar solutions in passive and dissipative systems and interpreted them as a novel type of soliton molecule (*22*, *23*). These localized solutions exhibit striking analogies with quantum mechanical trapping and spectral tunneling, as each spectrally distinct soliton creates an attractive potential well that traps the other and the transfer of energy between separate spectral regions becomes possible (*22*). Further, these molecules could also be of interest to support applications in advanced communications systems or spectroscopy (*22*, *24*). However, since these pulses require a specific type of hybrid dispersion relation, their experimental observation is difficult.

Here we report the generation of this new type of generalized dispersion soliton molecule from a dispersion managed mode-locked laser incorporating an intracavity pulse-shaper. This configuration, which was recently used to demonstrate the first laser operating in the PQS regime (*19*), allows for the tailoring of the net-cavity dispersion. The dispersion required for the generation of this new soliton bound state is achieved by applying a phase mask that is a combination of positive quadratic and negative quartic dispersion. Spectral, temporal and phase-resolved measurements confirm that the laser output pulses correspond to the type of solutions reported in previous numerical studies (*20*, *22*). The output spectrum exhibits two separated peaks, while the associated temporal shape displays strong interference fringes modulated by a hyperbolic secant envelope. We then expand on our interpretation and by adjusting the net-cavity dispersion profile we demonstrate the generation of a soliton molecule formed by a bound state of three spectral solitons. Our experimental results are in excellent agreement with theoretical predictions and realistic numerical simulations based on an iterative cavity map, and we point out a direct analogy between our experiments and conventional multislit diffraction experiments in the Fraunhofer regime. We expect our results to expand the scope of research on soliton molecules, which have been the focus of extensive work over the last decade due to the analogy with matter molecules (*25–27*), to bound states between spectrally distinct regimes. We also believe that this work will stimulate follow-up investigation of pulse dynamics and interaction in systems with complicated dispersion profiles. Finally, our platform offers a convenient playground to develop optical analogies to quantum mechanical phenomena such as spectral tunneling or mutual particle trapping.

To illustrate the concept of such polychromatic soliton molecules, we recall the conceptual approach reported by Akhmediev and Karlsson (*28*). Their description of a conventional soliton is shown schematically in Fig. 1A. The linear dispersion relation $\beta(\omega)$, expanded about a central frequency $\omega_0$, and in the frame moving at the associated group velocity, is denoted by the black curve. A soliton that is stationary in this frame is represented by a line that is horizontal (since the soliton is stationary in that frame) and straight (since solitons have a uniform phase), positioned above $\beta(\omega)$ by an amount $K\gamma P$, where $\gamma$ is the nonlinear parameter of the waveguide, P is the soliton's peak power and $K=½$ for conventional nonlinear Schrödinger solitons. This positioning ensures that the soliton does not intersect with the linear dispersion relation and therefore does not suffer from loss to linear waves. The pulse spectrum is concentrated around the peak in $\beta(\omega)$, as shown by the color gradient of the soliton line. The corresponding temporal profile of this soliton is shown in Fig. 1D.

In the presence of positive quadratic and negative quartic dispersion, the linear dispersion relation is double-peaked and symmetric about the central frequency $\omega_0$ (see Fig. 1B black curve). Because of the presence of higher-order dispersion, the magnitude and sign of the quadratic dispersion varies with wavelength. While the quadratic dispersion is positive at $\omega_0$, it is negative at each of the peaks centered at $\pm\omega_c$. With $\beta(\omega)$ locally parabolic at $\pm\omega_c$, solitons can form at the frequencies corresponding to each peak, producing a distinctive multi-peaked spectrum (*20*), as shown by the dashed orange curve in Fig. 1B. The frequency difference of these peaks is associated with the frequency of the temporal beat pattern (blue curve in Fig. 1E). The envelope of the soliton molecule (dashed orange curve Fig. 1E) satisfies the nonlinear Schrödinger equation with the relevant quadratic dispersion associated with each of the peaks in $\beta(\omega)$ (*20*). We can then generalize this concept to the bound states of more than two solitons. Figure 1C shows a triple-peaked linear dispersion relation (black curve) with peaks of equal negative curvature. Soliton atoms can form in each of these regions and their nonlinear binding forms a soliton molecule whose temporal intensity (blue curve in Fig. 1F) is again modulated by a hyperbolic secant envelope (dashed orange curve Fig. 1F).

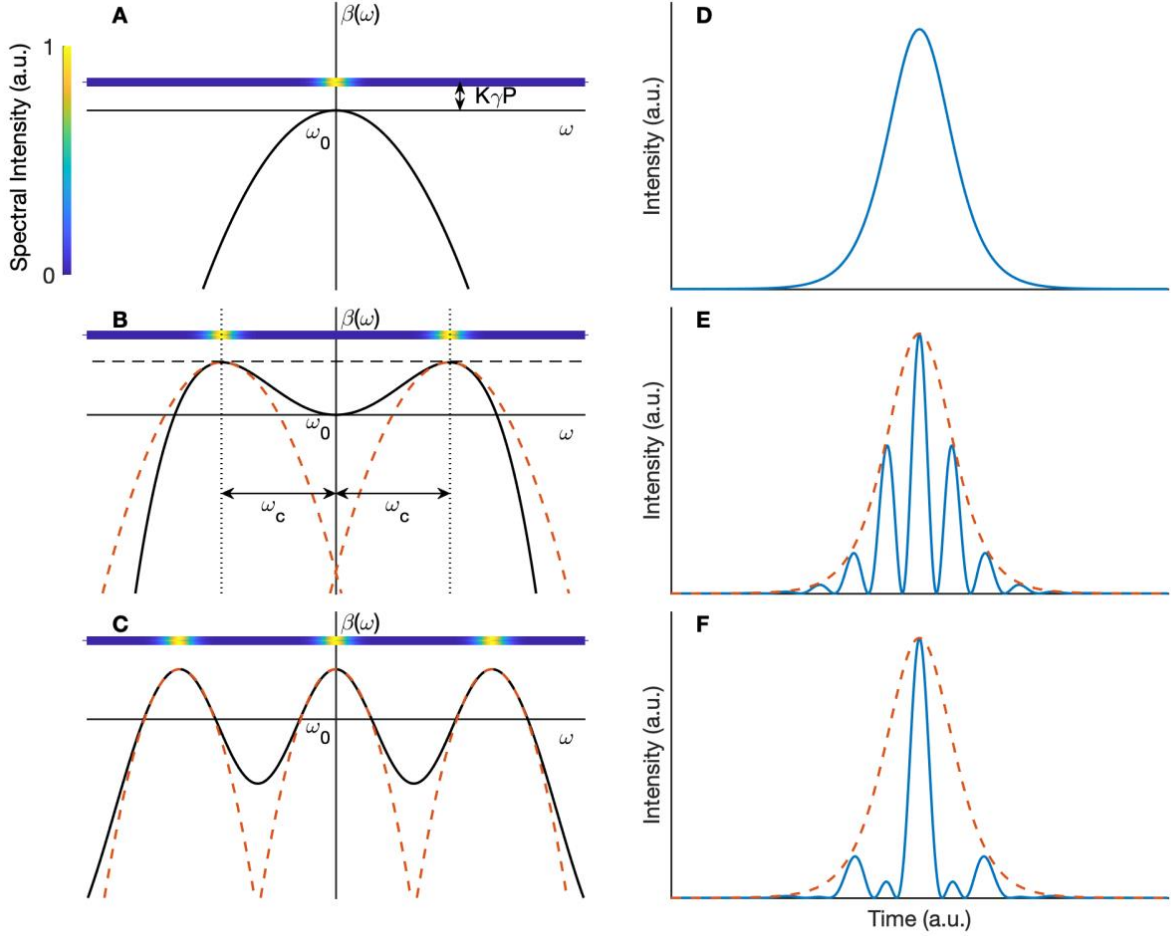

**Fig. 1. Hybrid linear dispersion relations and their corresponding solitons.** From top to bottom; single, double and triple-peaked dispersion relations. (**A-C**) Linear dispersion relation (black). Spectrum of soliton in rest frame sitting above $\beta(\omega)$ by an amount $\tfrac{1}{2}\gamma P$. The color gradient represents spectral intensity with increasing intensity going from blue to yellow. The central expansion frequency for the dispersion relations is $\omega_0$. When the pulse spectrum is localized around peaks in $\beta(\omega)$, each dispersion relation can be approximated by the dashed orange parabolic curves. (**D-F**) Temporal intensity of solitons shown in first row (blue) and their hyperbolic secant shaped envelopes (dashed orange).

In our experimental setup, we used a conventional mode-locked fiber laser incorporating a spectral pulse shaper (*19*) which is used to apply an arbitrary phase mask of the form

$$\phi(\omega) = L\left(\left(\frac{\beta_{2smf}(\omega-\omega_0)^2}{2} + \frac{\beta_{3smf}(\omega-\omega_0)^3}{6}\right) + \left(\frac{\beta_2(\omega-\omega_0)^2}{2} + \frac{\beta_4(\omega-\omega_0)^4}{24}\right) + \cdots\right). \quad (1)$$

The first term on the right-hand side of Eq. 1 corresponds to the phase compensating for the quadratic and cubic dispersion of the rest of the cavity which is dominated by single-mode fiber (SMF-28) segments. The cavity length is $L = 21.4$ m and the coefficients are fixed at $\beta_{2\text{smf}} = +21.4$ ps2/km and $\beta_{3\text{smf}} = -0.12$ ps3/km throughout this paper, based on values reported in ref. (*19*, *29*) for similar SMF. The second term accounts for the positive quadratic and negative quartic dispersion combination required to support the generation of the two-color soliton molecule. For the results discussed immediately below, we set $\beta_4 = -135$ ps4/km and we consider four different values of $\beta_2$.

The results of the spectral, temporal and phase-resolved measurements for each case are shown in the first four rows of Fig. 2, where each row corresponds to a different value of applied positive quadratic dispersion at $\omega_0$ (*30*). From the top to fourth row, $\beta_2 = 0$, 35, 50 and 70 ps2/km. For $\beta_2 = 0$ ps2/km the laser operates in the PQS regime (*19*). While the applied dispersion (orange curve in the left column) shows a single maximum for $\beta_2 = 0$ ps2/km, for nonzero, positive $\beta_2$ coefficients, the dispersion relation displays two maxima with a spacing that is proportional to $\sqrt{\beta_2}$. The blue curves in the left column show the corresponding measured output spectra. As the value of $\beta_2$ is increased, two distinct and identical maxima form in the dispersion relation and the spectra exhibit increasingly well-defined peaks centered at each of these maxima. For the largest applied positive quadratic dispersion, the spectrum comprises essentially two nearly independent peaks, as seen in Fig. 2D. We also note the presence of spectral sidebands corresponding to dispersive waves (*31*), and the spike at 1562 nm corresponding to a low-power continuous wave that does not affect our analysis.

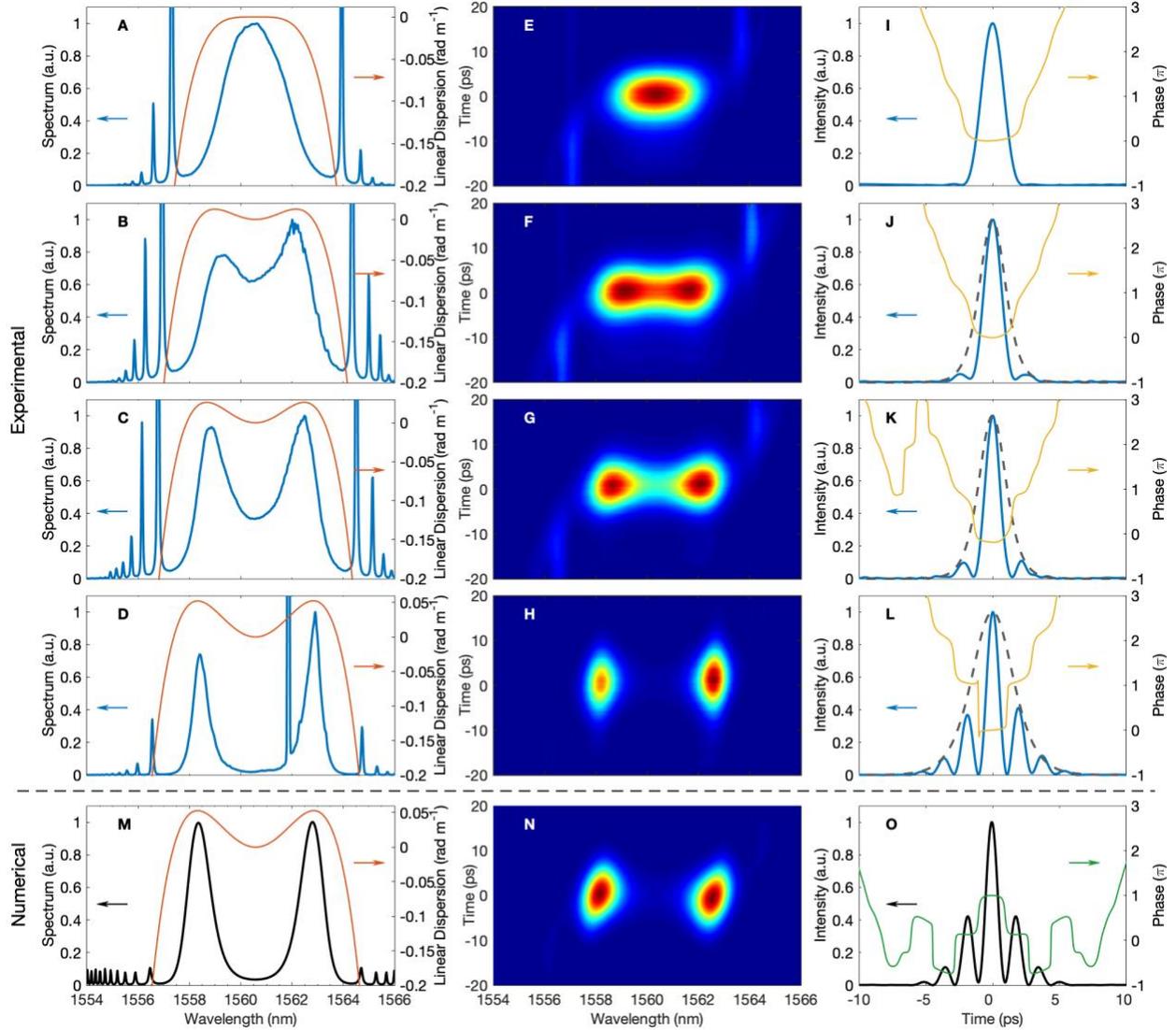

**Fig. 2. Measurements of two-colour soliton molecules.** The first four rows show experimental results. From top to fourth row, $\beta_2 = 0, 35, 50$ and $70$ ps2/km. The net cubic and quartic dispersion are fixed at $\beta_3 = 0$ ps3/km and $\beta_4 = -135$ ps4/km. The final row shows associated numerical results for $\beta_2 = 70$ ps2/km. (**A-D**) Measured output spectra (blue) and linear dispersion relation (orange). (**E-H**) Measured spectrograms. (**I-L**) Retrieved temporal intensity profiles (blue), temporal phase (yellow) and hyperbolic secant shaped meta-envelopes (dashed). (**M-O**) Numerical simulations with parameter values matching experimental values used in **D, H** and **L**. (**M**) Simulated output pulse spectrum (black) and linear dispersion relation (orange). (**N**) Simulated spectrogram. (**O**) Simulated temporal intensity (black) and temporal phase (green).

The middle column of Fig. 2 shows the spectrograms of the output pulses. These results demonstrate that the two pulses are temporally locked, even as the separation of the spectral peaks increases. Finally, the column on the right shows the temporal intensities (blue curve), and phases (yellow curve). As the value of $\beta_2$ increases, the pulses start to display more pronounced temporal fringes under an envelope that gradually broadens and has an approximate hyperbolic secant shape. As the temporal fringes develop, the width of the central maximum is reduced as it occupies less of the area under the meta-envelope. The full width at half maximum of the central peaks for the top to the fourth row are 1.89, 1.49, 1.25 and 1.03 ps. This reduction in pulse-width is consistent with the findings reported in ref. (*20*). The temporal fringe spacing $\Delta\tau$ is associated with the difference between the central frequencies of the soliton atoms, which are in turn determined by the positions of the peaks in the dispersion relation. These fringe spacings are given by $\Delta\tau = 2\pi/(2\omega_c)$ and the measured values for $\beta_2 = 35$, 50 and 70 ps$_2$/km are 2.50, 2.19 and 1.84 ps, respectively. These are in excellent agreement with the expected values of 2.57, 2.14 and 1.80 ps, respectively. Note that the π phase jumps in the retrieved temporal phase coincide with the intensity nodes, indicating that the phase of each of the solitons is constant and that the phases are mutually locked. These results provide strong evidence that we are observing soliton molecules with increasing spectral separation. The soliton atoms are associated with each of the peaks in the dispersion relation. These solitons travel at the same group velocity, are coherent and coincident; their nonlinear binding leads to the formation of a soliton molecule (*22*).

The cavity dynamics were modelled using an iterative cavity map (*19*). The propagation through each element was calculated by solving the generalized nonlinear Schrödinger equation with Kerr nonlinearity and dispersion up to the fourth order (*30*). The simulated spectrum, spectrogram and temporal profile, for simulation parameter values which match the experimental values in Fig. 2D, H and L (fourth row), are shown in the bottom row of Fig. 2 and are in very good agreement with the experimental results. In particular, the positions and relative amplitudes of the fringes in temporal intensity and the phase are all consistent with the measured results. (The complete set of numerical simulations for every experimental result in Fig. 2 is shown in Fig. S1).

To gain more insight into these bound states, we analyzed the constituent soliton atoms individually. While applying a phase mask with the same *β*$_2$ and *β*$_4$ parameters as for Fig. 2C, G, and K, (third row) one of the soliton pulses was filtered using the intracavity pulse-shaper. The measured output spectrum is shown in Fig. 3A (blue curve) and displays a single soliton centered at 1562.8 nm. This is confirmed by the corresponding spectrogram shown in the inset of Fig. 3A, in which the dashed white lines indicate the locations of the peaks and the central wavelength in the linear dispersion relation. As seen in Fig. 3B (blue curve), the temporal fringes have vanished, and the retrieved temporal intensity now has a hyperbolic secant shape. The temporal phase (yellow curve) is linear across the entire pulse width, indicating that the instantaneous frequency is constant. This linear phase arises from the offset between the single soliton and the reference frequency $\omega_0$. This offset of 1.97 nm corresponds to a difference in angular frequency of $1.52 \cdot 10^{12}$ rad s$^{-1}$. For the 4.4 ps span of the retrieved pulse, this leads to an expected phase change of 2.13π, in excellent agreement with the measured value of 2.08π. For comparison, we show the temporal shape (black dashed curve) when the short wavelength soliton is not filtered by the pulse shaper.

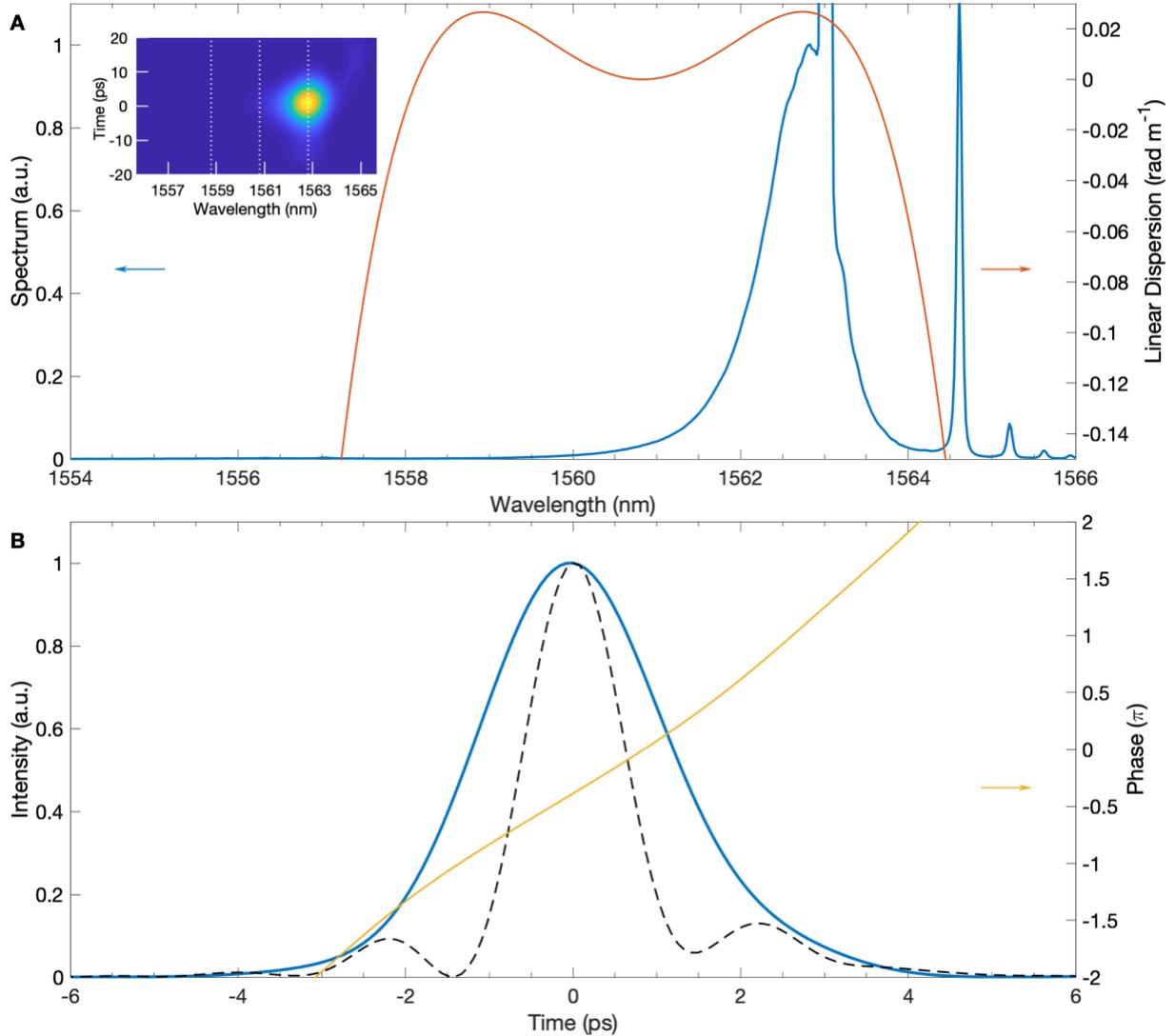

**Fig. 3. Analysis of individual soliton atoms of the molecule.** (**A**) Measured output spectrum (blue curve) and net linear dispersion relation (orange curve). The inset shows the measured spectrogram with white dashed lines denoting the locations of the peaks and central frequency in the dispersion relation. (**B**) Retrieved temporal intensity (blue curve) and temporal phase (yellow curve) of single-peaked spectrum and retrieved temporal intensity of double-peaked spectrum (dashed black curve).

These results suggest that the soliton bound states can be qualitatively understood as the spectral analogue of conventional double slit interference experiments in the Fraunhofer regime (*32*). In this analogy the common envelope, which in diffraction experiments is determined by the shape of the individual slits, is associated with the shape of the individual solitons. This is, in turn, determined by the curvature at each of the maxima in the dispersion relation, and the intensity. In interference experiments, the rapidly varying fringes arise from the spatial interference effects between the two slits, whereas here they arise from the temporal interference between the two solitons—the period of this interference pattern is determined by their mutual spacing $d$, and equals $2\pi/d$ in interference experiments, and $2\pi/(2\omega_c)$ in our experiments. Using this analogy, it is straightforward to understand the results in Fig. 3; when one of the slits is closed the interference disappears.

We now exploit the analogy with diffraction to generalize our interpretation to bound states of more than two solitons. Following the approach illustrated in Fig. 1C, we applied a phase mask so that the net-cavity linear dispersion has three evenly spaced peaks with equal negative curvature. Such a profile cannot be achieved using only quadratic and quartic dispersion, described by Eq. 1. Thus, we set $\beta_2 = \beta_4 = 0$ and applied a phase mask combining eighth order dispersion and a cosine function (*30*). The resulting applied linear dispersion profile is shown in Fig. 4A (orange curve). The associated measured output spectrum (blue curve) displays a soliton centered at each of the three peaks in the dispersion relation.

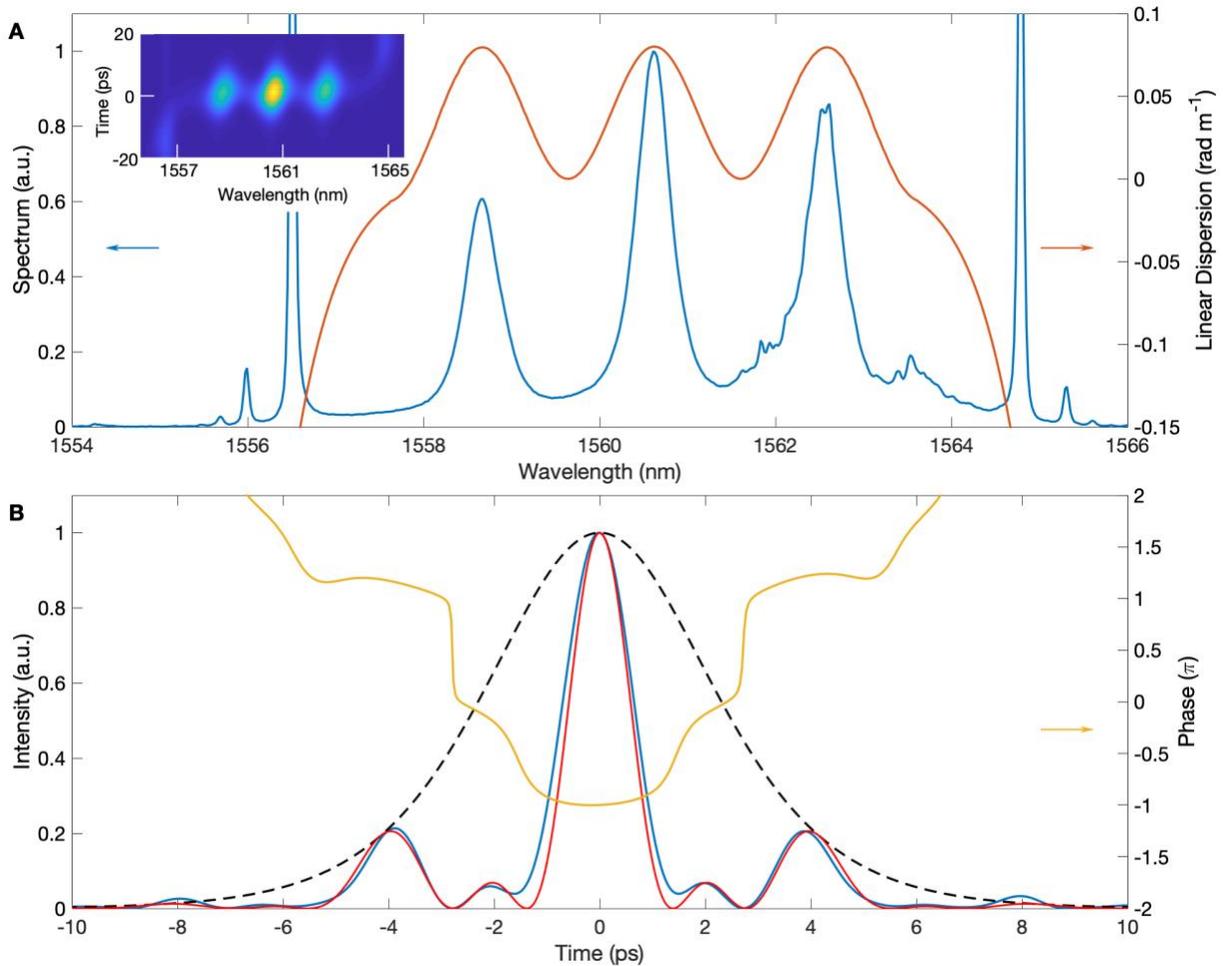

**Fig. 4. Three color soliton molecule.** (**A**) Measured output spectrum (blue curve) and applied net-cavity linear dispersion relation (orange curve). The inset shows the measured spectrogram. (**B**) Retrieved temporal intensity (blue curve), predicted temporal intensity (red curve), hyperbolic secant shaped envelope (dashed black curve) and retrieved temporal phase (yellow curve).

The retrieved temporal intensity is described by the blue curve in Fig. 4B and displays primary and secondary maxima in good agreement with the predicted shape (red curve) illustrated in Fig.

1F, and which is similar to that of a three-slit diffraction experiment. Based on the spectral separation between the peaks of the dispersion profile, the expected fringe period for the primary and secondary maxima are 3.77 ps and 1.89 ps, respectively. By modulating this double period interference pattern with a hyperbolic secant envelope (black dashed curve) that has the same width as the envelope of the measured temporal intensity (full width at half maximum of 5 ps) we arrive at a function that corresponds to the predicted temporal shape and is shown in Fig. 4B (red curve). The agreement between the measured and predicted temporal shapes is remarkable both for the positions of the maxima and their relative amplitudes, up to the third primary maximum. We also note that the retrieved temporal phase is flat across the entire pulse width, aside from sign changes, confirming that the three pulses correspond to solitons that are mutually coherent.

In conclusion, we have reported the first experimental observations of polychromatic soliton molecules. These solitons have novel temporal intensities and pulse spectra and we show that they can be thought of as bound states of multiple conventional nonlinear Schrödinger solitons which propagate with identical group velocities. We demonstrated that by removing one of the atoms in a two-color soliton molecule, the temporal fringes vanish, and a conventional soliton is recovered. We also reported the formation of a polychromatic soliton molecule formed by the nonlinear binding between three solitons. These measurements demonstrate that regions of local negative curvature are sufficient for soliton formation even when $\beta_2$ is positive at other frequencies. Our observations not only confirm earlier theoretical and numerical work (*20*, *22*, *23*), but extend the scope of this work considerably.

It is in principle possible to produce a polychromatic soliton molecule in a waveguide with fixed dispersion (*33*, *34*). However, the design of such waveguides is very challenging, and the fabrication tolerances would likely be exceedingly tight. Thus, a crucial innovation presented here is the use of the intracavity pulse shaper to introduce complicated hybrid dispersion profiles. The ability to vary the net dispersion of the cavity, and thus to vary the pulse spectra, was crucial to concentrating power around the peaks in $\beta(\omega)$. This "lumped-element" approach does, however, result in the shape of the pulse changing throughout the cavity since all the desired dispersion characteristics are applied in a single cavity element. This, in addition to the insertion losses, bandwidth and spectral resolution of the pulse shaper, limits the beat frequency and the size of the secondary temporal maxima. These limitations could be alleviated by using a pulse shaper with lower insertion losses, larger bandwidth and higher resolution. On the other hand, the fact that solitons molecules can be observed in our present geometry highlights their stability to significant perturbations.

As the number of soliton atoms $M$ grows, the central temporal peak would become increasingly dominant—its peak power would grow as $M^2$ and its width would scale as $M^{-1}$. In the limit $M \to \infty$ the response would be the temporal equivalent of a grating. In addition to increasing $M$, Fig. 2 demonstrates that the central maximum of the temporal intensity also becomes narrower with increasing $\beta_2$. This is consistent with ref. (*20*) and suggests that the inclusion of moderate positive $\beta_2$ could further improve the energy-width scaling of optical pulses for laser applications (*16*, *19*). The exquisite control over the net-cavity dispersion allowed by our setup offers an ideal platform to investigate novel molecular-like structures formed by multiple atoms in nonlinear systems. Such complex polychromatic solitons could provide new insights into soliton physics and other fields including molecular biology or quantum mechanics (*35*).

**References**

1. K. E. Lonngren, Soliton experiments in plasmas. *Plasma Phys.* **25**, 943–982 (1983).
2. E. Polturak, P. G. N. Devegvar, E. K. Zeise, D. M. Lee, Solitonlike propagation of zero sound in superfluid 3He. *Phys. Rev. Lett.* **46**, 1588–1591 (1981).
3. I. P. Christov, M. M. Murnane, H. C. Kapteyn, J. Zhou, C.-P. Huang, Fourth-order dispersion-limited solitary pulses. *Opt. Lett.* **19**, 1465–1467 (1994).
4. J. Denschlag, Generating solitons by phase engineering of a Bose-Einstein condensate. *Science.* **287**, 97–101 (2000).
5. I. Jung, F. Kärtner, N. Matuschek, Self-starting 6.5-fs pulses from a Ti: sapphire laser. *Opt. Lett.* **22**, 1009–1011 (1997).
6. L. F. Mollenauer, R. H. Stolen, The soliton laser. *Opt. Lett.* **9**, 13–15 (1984).
7. J. M. Dudley, G. Genty, S. Coen, Supercontinuum generation in photonic crystal fiber. *Rev. Mod. Phys.* **78**, 1135–1184 (2006).
8. T. Herr, V. Brasch, J. D. Jost, C. Y. Wang, N. M. Kondratiev, M. L. Gorodetsky, T. J. Kippenberg, Temporal solitons in optical microresonators. *Nat. Photonics.* **8**, 145–152 (2014).
9. H. Bao, A. Cooper, M. Rowley, L. Di Lauro, J. S. Totero Gongora, S. T. Chu, B. E. Little, G. L. Oppo, R. Morandotti, D. J. Moss, B. Wetzel, M. Peccianti, A. Pasquazi, Laser cavity-soliton microcombs. *Nat. Photonics.* **13**, 384–389 (2019).
10. T. Brabec, S. M. J. Kelly, Third-order dispersion as a limiting factor to mode-locking in femtosecond solitary lasers. *Opt. Lett.* **18**, 2002–2004 (2002).
11. A. Höök, M. Karlsson, Ultrashort solitons at the minimum-dispersion wavelength: effects of fourth-order dispersion. *Opt. Lett.* **18**, 1388–1390 (1993).
12. Y. Kodama, M. Romagnoli, M. Midrio, S. Wabnitz, Role of third-order dispersion on soliton instabilities and interactions in optical fibers. *Opt. Lett.* **19**, 165–167 (1994).
13. M. L. Dennis, I. N. Duling, Third-order dispersion in femtosecond fiber lasers. *Opt. Lett.* **19**, 1750–1752 (1994).
14. S. Roy, F. Biancalana, Formation of quartic solitons and a localized continuum in silicon-based slot waveguides. *Phys. Rev. A* **87**, 025801 (2013).
15. A. Blanco-Redondo, C. M. De Sterke, J. E. Sipe, T. F. Krauss, B. J. Eggleton, C. Husko, Pure-quartic solitons. *Nat. Commun.* **7**, 10427 (2016).
16. K. K. K. Tam, T. J. Alexander, A. Blanco-Redondo, C. Martijn de Sterke, Stationary and dynamical properties of pure-quartic solitons. *Opt. Lett.* **44**, 3306–3309 (2019).
17. H. Taheri, A. B. Matsko, Quartic dissipative solitons in optical Kerr cavities. *Opt. Lett.* **44**, 3086–3089 (2019).
18. C.-W. Lo, A. Stefani, C. M. de Sterke, A. Blanco-Redondo, Analysis and design of fibers for pure-quartic solitons. *Opt. Express.* **26**, 7786–7796 (2018).

19. A. F. J. Runge, D. D. Hudson, K. K. K. Tam, C. M. de Sterke, A. Blanco-Redondo, The pure-quartic soliton laser. *Nat. Photonics* (2020), doi:https://doi.org/10.1038/s41566-020-0629-6.

20. K. K. K. Tam, T. J. Alexander, A. Blanco-Redondo, C. M. De Sterke, Generalized dispersion Kerr solitons. *Phys. Rev. A*. **101**, 043822 (2020).

21. V. I. Kruglov, J. D. Harvey, Solitary waves in optical fibers governed by higher-order dispersion. *Phys. Rev. A*. **98**, 063811 (2018).

22. O. Melchert, S. Willms, S. Bose, A. Yulin, B. Roth, F. Mitschke, U. Morgner, I. Babushkin, A. Demircan, Soliton Molecules with Two Frequencies. *Phys. Rev. Lett.* **123**, 243905 (2019).

23. O. Melchert, A. Yulin, A. Demircan, Dynamics of localized dissipative structures in a generalized Lugiato-Lefever model with negative quartic group-velocity dispersion. *Opt. Lett.* **45**, 2764–2767 (2020).

24. P. Rohrmann, A. Hause, F. Mitschke, Solitons beyond binary: Possibility of fibre-optic transmission of two bits per clock period. *Sci. Rep.* **2**, 866 (2012).

25. M. Stratmann, T. Pagel, F. Mitschke, Experimental observation of temporal soliton molecules. *Phys. Rev. Lett.* **95**, 143902 (2005).

26. G. Herink, F. Kurtz, B. Jalali, D. R. Solli, C. Ropers, Real-time spectral interferometry probes the internal dynamics of femtosecond soliton molecules. *Science*. **356**, 50–54 (2017).

27. Z. Q. Wang, K. Nithyanandan, A. Coillet, P. Tchofo-Dinda, P. Grelu, Optical soliton molecular complexes in a passively mode-locked fibre laser. *Nat. Commun.* **10**, 830 (2019).

28. N. Akhmediev, M. Karlsson, Cherenkov radiation emitted by solitons in optical fibers. *Phys. Rev. A*. **51**, 2602–2607 (1995).

29. K. Hammani, B. Kibler, C. Finot, P. Morin, J. Fatome, J. M. Dudley, G. Millot, Peregrine soliton generation and breakup in standard telecommunications fiber. *Opt. Lett.* **36**, 112–114 (2011).

30. Materials and methods are available as supplementary materials at the Science website.

31. S. M. J. Kelly, Characteristic sideband instability of periodically amplified average soliton. *Electron. Lett.* **28**, 806–807 (1992).

32. F. A. Jenkins, H. E. White, *Fundamentals of Optics* (McGraw Hill, 1957).

33. L. Zhang, Q. Lin, Y. Yue, Y. Yan, R. G. Beausoleil, A. E. Willner, Silicon waveguide with four zero-dispersion wavelengths and its application in on-chip octave-spanning supercontinuum generation. *Opt. Express*. **20**, 1685–1690 (2012).

34. W. H. Reeves, D. V. Skryabin, F. Biancalana, J. C. Knight, P. S. J. Russell, F. G. Omenetto, A. Efimov, A. J. Taylor, Transformation and control of ultra-short pulses in dispersion-engineered photonic crystal fibres. *Nature*. **424**, 511–515 (2003).

35. T. Dauxois, M. Peyrard, *Physics of Solitons* (Cambridge University Press, 2006).

36. C. Dorrer, I. Kang, Simultaneous temporal characterization of telecommunication optical pulses and modulators by use of spectrograms. *Opt. Lett.* **27**, 1315–1317 (2002).

37. R. Trebino, *Frequency-resolved optical gating: the measurement of ultrashort laser pulses* (Springer, 2000).

38. B. Oktem, C. Ülgüdür, F. Ö. Ilday, Soliton-similariton fibre laser. *Nat. Photonics*. **4**, 307–311 (2010).

39. G. P. Agrawal, *Nonlinear fiber optics* (Academic Press, 1995).

**Acknowledgments**

**General:** The authors thank Mr. Kevin K. K. Tam for fruitful discussions.
**Funding:** This work was supported by the Australian Research Council (ARC) Discovery Project (grant no. DP180102234), the University of Sydney Professor Harry Messel Research Fellowship and the Asian Office of Aerospace R&D (AOARD) (grant no FA2386-19-1-4067).
**Author contributions:** A.F.J.R., D.D.H. and A.B.-R. designed the experiment. J.P.L. performed the experiments and the numerical simulations. J.P.L., A.F.J.R., T.J.A and C.M.d.S carried out the theoretical analysis. C.M.d.S. supervised the overall project. All the authors contributed to interpretation of the data and wrote the manuscript.
**Competing interests:** The authors declare no conflicts of interest.
**Data and materials availability:** The data that support the plots in this paper and other finding of this study are available from the corresponding author on reasonable request.

# Supplementary information for Polychromatic soliton molecules

Joshua P. Lourdesamy[1], Antoine F. J. Runge[1]*, Tristram J. Alexander[1], Darren D. Hudson[2], Andrea Blanco-Redondo[3], C. Martijn de Sterke[1,4]
*Corresponding author. Email: antoine.runge@sydney.edu.au

Experiments: Cavity configuration

A schematic diagram of the laser cavity used in our experiments is shown in Fig. S1. The laser is an $L = 21.44$ m long erbium-doped fiber laser operating around 1560 nm that uses nonlinear polarization evolution for mode-locking. The pulse shaper is based on a spatial light modulator (Finisar WaveShaper 4000S) that produces an arbitrary phase mask and is used to adjust the net-cavity dispersion (*19*).

Two laser diodes producing light at 980 nm were used to couple light into the cavity through two 980/1550 nm wavelength division multiplexers. An optical isolator ensures that light can only propagate in one direction within the cavity. Passive mode-locking is achieved using two fiber polarization controllers and a fiber polarizer to serve as an artificial saturable absorber.

At a pump power of approximately 500 mW, the laser is self-starting and multi-pulsing after adjusting the polarization controller. Single pulsing is achieved by reducing the pump power to approximately 180 mW.

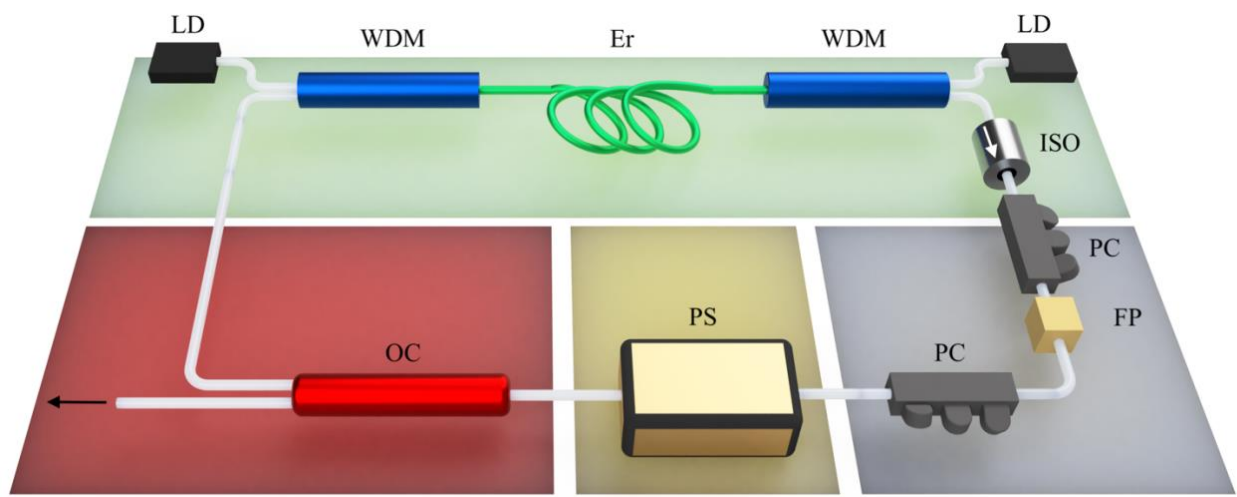

**Fig. S1. Laser cavity for generation of polychromatic soliton molecules.** LD, laser diode; ISO, optical isolator; PC, polarization controller; FP, in-line fiber polarizer; PS, pulse shaper; OC, output coupler; WDM, wavelength division multiplexers; Er, erbium-doped fiber.

Phase-resolved characterization method

The cavity output was coupled into the FREG apparatus (*15, 36*). The pulses were split into two branches using a 70/30 fiber-coupler. The 30% is passed through a variable delay and then to a fast photodiode which detects and transfers the pulse to the electronic domain. This electronic signal was then used to drive a Mach-Zehnder modulator which gated the power from the branch with 70% of the output power. Using an optical spectrum analyzer (OSA), we measured the spectra as a function of delay in order to produce an optical spectrogram for each pulse. We then numerically de-convolve these spectrograms (512x512 grid-retrieval errors < 0.003) to retrieve the temporal pulse intensity and phase (*37*). We then verified the validity of the retrieved pulses by reconstructing their spectrograms and by comparing their Fourier transforms to their spectra measured with the OSA.

Numerical simulation model

Numerical simulations are based on the nonlinear Schrödinger equation (NLSE):

$$\frac{\partial A}{\partial z} = -\frac{\alpha_l}{2}A - i\frac{\beta_2}{2}\frac{\partial^2 A}{\partial t^2} + \frac{\beta_3}{6}\frac{\partial^3 A}{\partial t^3} + i\frac{\beta_4}{24}\frac{\partial^4 A}{\partial t^4} + \frac{g}{2}A + i\gamma A|A|^2. \tag{1}$$

Where *A(z, t)* is the slowly varying amplitude of the pulse, *z* is the propagation coordinate, *t* is the time in the frame of the pulse, $\alpha_l$ is the linear loss, $\gamma$ is the nonlinear parameter given by $\gamma = n_2\omega_0/(cA_{eff})$, where $n_2$ is the nonlinear refractive index, $\omega_0$ the central frequency, *c* the speed of light in vacuum and $A_{eff}$ the effective mode area.

The gain *g* in the doped fiber section (and zero in all other sections of the cavity) is calculated using:

$$g = \frac{g_0}{1 + E(z)/E_{sat}}, \tag{2}$$

where $g_0 = 3.45$ is the small-signal gain, *E(z)* is the pulse energy and $E_{sat}$ is the saturation energy. We multiply *g(z)* with a Lorentzian profile of 50 nm width to form the finite gain bandwidth $g(z,\omega_0)$ (*38*). The saturable absorber is modelled by a transfer function that describes its power-dependent transmittance

$$T(r) = 1 - \frac{q_0}{1 + P/P_0}, \tag{3}$$

where $q_0$ is the unsaturated loss of the saturable absorber, $P(\tau)=|A(z,\tau)|^2$ is the instantaneous pulse power and $P_0$ is the saturation power. The spectral pulse shaper is modelled by multiplying the electric field in the spectral domain by the following function

$$A(\omega) = A_{in}(\omega)e^{i\phi(\omega)L}, \tag{4}$$

Where $A_{in}(\omega)$ is the amplitude modulation and $\phi(\omega)$ is the applied phase profile and $L$ is the cavity optical path length. The insertion losses (~5.6 dB) of the spectral pulse shaper are also considered in the simulations.

Our numerical model is solved with a standard symmetric split-step Fourier method algorithm (*39*). The dispersion and gain contributions are calculated in the frequency domain, whereas the nonlinear term is calculated in the time domain. For our simulations we have used an initial field composed of Gaussian random noise multiplied by a sech shape in the time domain. The same stable solutions are reached for different initial noise fields.

The values of the simulation parameters are similar to the experimental values. Most of the cavity is comprised of SMF-28 fiber which has an $\alpha_l = 0.44$ μm-1, a mode-filed diameter (MDF) of 10.4 μm, a numerical aperture (NA) of 0.14 and $\gamma = 0.0013$ W-1m-1 at 1560.5 nm. The erbium-doped fiber used has $L = 1.5$ m, MDF = 9.5 μm, NA = 0.13 and $\gamma = 0.0016$ W-1m-1. As the NA and MDF for the SMF and erbium-doped fiber segments are almost identical, we assume that the dispersion coefficients for both are $\beta_2 = -21.4$ ps2km-1, $\beta_3 = 0.12$ ps3km-1 and $\beta_4 = -0.0022$ ps4km-1.

The simulated output pulse characteristics for the experimental cases presented in Fig. 2 (see main text) are shown in Fig. S2. The first column shows the pulse spectra. In all cases the simulated pulses (red curves) are in very good agreement with the experimental results (black curves). The second column shows the simulated spectrograms, these are in good agreement with those shown in Fig. 2. The final column shows the temporal profiles of the pulses. The experimental (black curves) and the simulated (red curves) intensities are in excellent agreement, with both the positions and amplitudes of the fringes matching in all cases. The measured (green curves) and the simulated (dashed magenta) phase also agree well, with the differences attributed to arbitrary differences in direction of the $\pi$ phase flips that correspond to the intensity nodes.

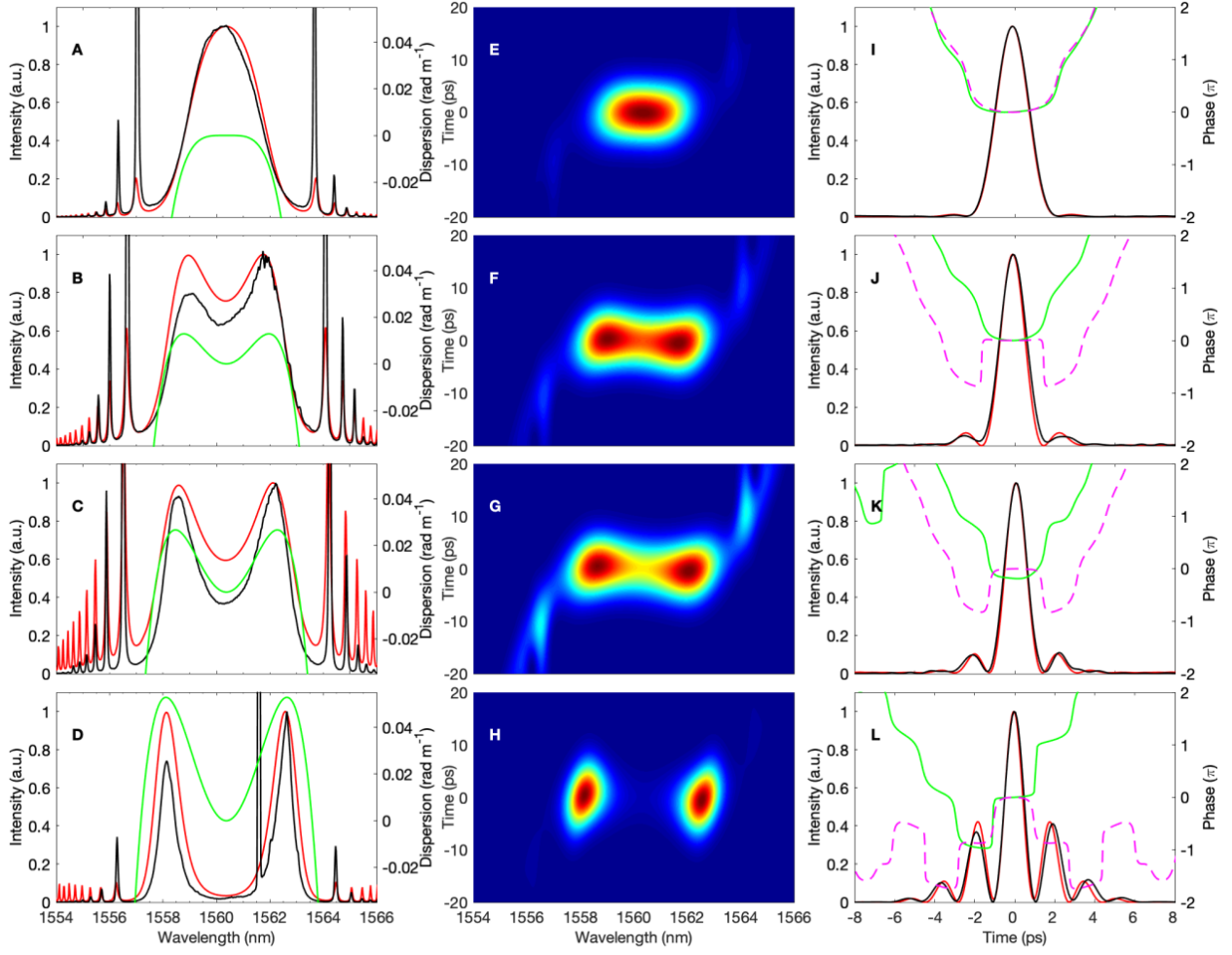

**Fig. S2. Simulated output pulse characteristics.** From top to bottom row, $\beta_2 = 0, 35, 50$ and $70$ ps2/km. The net cubic and quartic dispersion are fixed at $\beta_3 = 0$ ps3/km and $\beta_4 = -135$ ps4/km. (**A-D**) Measured output spectra (black), simulated output spectra (red) and linear dispersion relation (green). (**E-H**) Simulated spectrograms. (**I-L**) Retrieved temporal intensity (black) and phase (green), simulated temporal intensity (red) phase (dashed magenta).

Three color soliton molecule dispersion profile

The phase mask applied in order to generate the net-cavity linear dispersion relation required to produce the three color soliton molecule (see Fig. 4) is described by

$$\phi(\omega) = \left\{ L \left[ \left( \frac{\beta_{2smf}\omega^2}{2!} + \frac{\beta_{3smf}\omega^3}{3!} \right) + \left( \frac{\beta_8\omega^8}{8!} + 0.04 \left( \cos\frac{2\pi\omega}{1.5 \cdot 10^{12}} + 1 \right) \right) \right] \right\} \quad (5)$$

for $-2.3 \cdot 10^{12}\ rad/s\ <\ \omega < 2.3 \cdot 10^{12}\ rad/s$ and

$$\phi(\omega) = \left\{ L \left[ \left( \frac{\beta_{2smf}\omega^2}{2!} + \frac{\beta_{3smf}\omega^3}{3!} \right) + \left( \frac{\beta_8\omega^8}{8!} \right) \right] \right\} \quad (6)$$

elsewhere.

The first term on the right-hand side of Eq. 5 and 6 corresponds to the phase compensating for the native quadratic and cubic dispersion of the cavity, excluding the pulse shaper, which is primarily composed of single-mode fiber (SMF) segments. $L$ is the length of the cavity, $\beta_8 = -6.7$ ps8/km, $\beta_{2smf} = +21.4$ ps2/km and $\beta_{3smf} = -0.12$ ps3/km, based on values reported in ref. (*19*, *29*) for similar SMF.